\documentclass[aps,prb,twocolumn,superscriptaddress,showpacs,10pt]{revtex4-1}
\usepackage{graphicx}
\usepackage{hyperref}
\begin{document}

% Use the \preprint command to place your local institutional report
% number in the upper righthand corner of the title page in preprint mode.
% Multiple \preprint commands are allowed.
% Use the 'preprintnumbers' class option to override journal defaults
% to display numbers if necessary
%\preprint{}

%Title of paper
\title{Direct Detection of Photoinduced Charge Transfer Complexes in Polymer:Fullerene Blends}

% repeat the \author .. \affiliation  etc. as needed
% \email, \thanks, \homepage, \altaffiliation all apply to the current
% author. Explanatory text should go in the []'s, actual e-mail
% address or url should go in the {}'s for \email and \homepage.
% Please use the appropriate macro foreach each type of information

% \affiliation command applies to all authors since the last
% \affiliation command. The \affiliation command should follow the
% other information
% \affiliation can be followed by \email, \homepage, \thanks as well.
\author{Jan Behrends}
\email[]{j.behrends@fu-berlin.de}
%\homepage[]{Your web page}
%\thanks{}
%\Altaffiliation{}
\affiliation{Fachbereich Physik, Freie Universit\"at Berlin, Arnimallee 14, D-14195 Berlin, Germany}
\author{Andreas Sperlich}
\affiliation{Experimental Physics VI, Julius-Maximilians Universit\"at W\"urzburg, Am Hubland, D-97074 W\"urzburg, Germany}
\author{Alexander Schnegg}
\affiliation{Institut f\"ur Silizium-Photovoltaik, Helmholtz-Zentrum Berlin f\"ur Materialien und Energie, Kekul\'estr.\ 5, D-12489 Berlin, Germany}
\author{Till Biskup}
\altaffiliation[Present address: ]{Physical and Theoretical Chemistry Laboratory, University of Oxford, UK}
\affiliation{Fachbereich Physik, Freie Universit\"at Berlin, Arnimallee 14, D-14195 Berlin, Germany}
\author{Christian Teutloff}
\affiliation{Fachbereich Physik, Freie Universit\"at Berlin, Arnimallee 14, D-14195 Berlin, Germany}
\author{Klaus Lips}
\affiliation{Institut f\"ur Silizium-Photovoltaik, Helmholtz-Zentrum Berlin f\"ur Materialien und Energie, Kekul\'estr.\ 5, D-12489 Berlin, Germany}
\author{Vladimir Dyakonov}
\affiliation{Experimental Physics VI, Julius-Maximilians Universit\"at W\"urzburg, Am Hubland, D-97074 W\"urzburg, Germany}
\affiliation{ZAE Bayern, Am Hubland, D-97074 W\"urzburg, Germany}
\author{Robert Bittl}
\affiliation{Fachbereich Physik, Freie Universit\"at Berlin, Arnimallee 14, D-14195 Berlin, Germany}
%Collaboration name if desired (requires use of superscriptaddress
%option in \documentclass). \noaffiliation is required (may also be
%used with the \author command).
%\collaboration can be followed by \email, \homepage, \thanks as well.
%\collaboration{}
%\noaffiliation

\date{\today}

\begin{abstract}
We report transient electron paramagnetic resonance (trEPR) measurements with sub-microsecond time resolution performed on a polymer:fullerene blend consisting of poly(3-hexylthiophene) (P3HT) and [6,6]-phenyl ${\rm C_{61}}$-butyric acid methyl ester (PCBM) at low temperature. The trEPR spectrum immediately following photoexcitation reveals signatures of spin-correlated polaron pairs. The pair partners (positive polarons in P3HT and negative polarons in PCBM) can be identified by their characteristic $g$-values. The fact that the polaron pair states exhibit strong non-Boltzmann population unambiguously shows that the constituents of each pair are geminate, i.e.\ originate from one exciton. We demonstrate that coupled polaron pairs are present even several microseconds after charge transfer and suggest that they embody the intermediate charge transfer complexes which form at the donor/acceptor interface and mediate the conversion from excitons into free charge carriers.
\end{abstract}

% insert suggested PACS numbers in braces on next line
\pacs{72.20.Jv, 87.80.Lg, 72.40.+w, 88.40.jr, 72.80.Le, 81.05.Fb, 76.30.Pk, 73.50.Pz}
% insert suggested keywords - APS authors don't need to do this
%\keywords{}

%\maketitle must follow title, authors, abstract, \pacs, and \keywords
\maketitle
% body of paper here - Use proper section commands
% References should be done using the \cite, \ref, and \label commands
%\section{Introduction}
% Put \label in argument of \section for cross-referencing
%\section{\label{}}
%
\section{Introduction}
Photoinduced charge transfer and subsequent charge separation are the key processes in organic bulk heterojunction solar cells. Upon light absorption in the polymer an exciton is created which dissociates at the heterojunction into a positive polaron (P\textsuperscript{+}) in the polymer and a negative polaron (P\textsuperscript{-}) in the electron accepting material.\cite{sariciftci1992_science} Several studies reported the charge transfer to occur within less than $1~{\rm ps}$ in a variety of polymer:fullerene blends.\cite{moses2000_cpl,brabec2001_cpl} The fast charge transfer effectively prevents the primary exciton from recombination and thus establishes the basis for quantum efficiencies close to unity\cite{park2009_natphot} and potentially efficient plastic solar cells.\cite{green2011_proginphot,deibel2010_repprogphys} Charge transfer is succeeded by the formation of an intermediate charge transfer complex (CTC) at the heterojunction,\cite{hasharoni1997_jchemphys} whose dissociation finally yields separated polarons. CTC dissociation times possibly exceed the charge transfer time by several orders of magnitude,\cite{mueller2005_prb} and the dissociation efficiency critically influences the photocurrent in polymer:fullerene solar cells.\cite{mihailetchi2004_prl} 

It is generally assumed that CTC resemble hybrid states at the heterojunction, formed by polymer HOMO and fullerene LUMO levels, occupied by the P\textsuperscript{+} and the P\textsuperscript{-}, respectively. Yet, strong discrepancies exist regarding the reported time constants. While time-resolved photoluminescence measurements on a blend of a fluorene copolymer and PCBM revealed CTC decay times as short as $4~{\rm ns}$\cite{veldman2008_jacs} modeling of $I$-$V$ characteristics requires CTC dissociation in the microsecond range to achieve agreement with experimental data.\cite{mihailetchi2004_prl,mihailetchi2005_prl} Recent transient absorption and microwave photoconductance measurements suggest that this process is very efficient, although the charge generation occurs via CTC, so that the yield is close to unity and almost temperature independent for annealed P3HT:PCBM.\cite{grzegorczyk2010_jphyschemc} Also ``hot processes'' are being discussed, leading to faster CTC dissociation due to excess energy from the donor excited state.\cite{lee2010_jacs,ohkita2008_jacs,clarke2010_advmat} Several studies utilizing optical techniques such as photoluminescence,\cite{hasharoni1997_jchemphys,hallermann2008_apl} photothermal deflection spectroscopy\cite{bensonsmith2007_afm} or electroluminescence\cite{tvingstedt2009_jacs} provided conclusive evidence for the existence of CTC by detecting either an emission band red-shifted to the polymer emission or a CTC absorption band energetically separated from the polymer absorption. However, since optical transitions often overlap spectrally, all-optical techniques lack microscopic selectivity, i.e.\ they cannot specifically address either P\textsuperscript{+} in the polymer or P\textsuperscript{-} in the electron accepting material. In consequence, the exact microscopic identity of CTC has remained speculative to date and the electronic states involved in CTC are subject to ongoing discussion.\cite {mihailetchi2004_prl,veldman2008_jacs,lee2010_jacs,drori2010_prb,deibel2010_advmat} In particular, experimental techniques providing direct access to the CTC dynamics are needed to finally solve this important issue.

\section{EPR of Polymer:Fullerene blends}
Light-induced electron paramagnetic resonance (EPR) spectroscopy has previously proven successful in identifying the existence of free polarons in different polymers and fullerene-based materials.\cite{sariciftci1992_science,dyakonov1999_prb} Owing to the microscopic selectivity for the distinct differences between the EPR fingerprints of polarons in polymers and fullerenes, EPR is commonly employed to verify charge transfer in various polymer:fullerene blends as well as mixtures between polymers and (inorganic) nanoparticles.\cite{deceuster2001_prb,keeble2009_jphyschemc,dietmueller2009_apl} The ability to resolve small interaction energies renders EPR appropriate to investigate weakly-coupled charge-carrier pairs. The typical time resolution of a conventional continuous-wave EPR (cwEPR) experiment employing field modulation and lock-in detection with full spectral resolution is in the range of milliseconds. It is thus possible to attain information on the dynamics of long-lived paramagnetic states by switching the illumination on and off and simultaneously detecting EPR.\cite{heinemann2009_afm} However, light-induced cwEPR lacks the time resolution that is required to detect intermediate CTC with lifetimes in the ${\rm ns}$ to ${\rm \mu s}$ range. Transient EPR (trEPR) following a laser flash omitting field modulation and lock-in detection, which is employed routinely to study charge transfer processes in biological systems,\cite{bittl2005_biochimica,lubitz2002_accchemres} allows studying charge carrier dynamics with sub-${\rm \mu s}$ time resolution.\cite{kim1976_jmr} This technique takes advantage of the fact that a non-Boltzmann polarization exists immediately after exciton dissociation, enhancing the EPR signal intensity and providing insight into the generation of charge-separated states. In particular, the geometry of (intermediate) contributing paramagnetic states and their dynamics can be accessed. Transient EPR was previously measured on oligothiophene:fullerene mixtures at room temperature, but no indication of CTC could be found.\cite{scharber1999_synthmet} Other trEPR studies on polythiophene:fullerene blends at several temperatures revealed clear signatures of spin-correlated charge carrier pairs.\cite{pasimeni2001_jmc,pasimeni2001_chemphys,franco2005_jphyschemb} The authors deduced a pair distance of $20-30~{\rm \AA}$ based on the interaction strength. They observed a change of the spectrum with increasing time after the laser flash and found significant variations of the spectrum when changing the temperature. Since no separated polarons were detected in that study, the role of the spin-correlated pairs in the course of free charge-carrier generation remained elusive. Yet, an in-depth understanding of the conversion process from coupled charge carrier pairs into separated polarons is crucial for understanding of photocurrent generation in organic solar cells

In this communication we report trEPR measurements with sub-$\rm \mu s$ time resolution performed on a polymer:fullerene blend at low temperatures. We show that the trEPR spectrum immediately following photoexcitation reveals signatures of spin-correlated polaron pairs (PP) --- as observed previously\cite{franco2005_jphyschemb} --- and decisively differs from the spectrum of separated polarons commonly observed in light-induced cwEPR. The PP partners exhibit spin-spin coupling, which can provide information on the distance within the pair. The fact that the PP spin states exhibit strong non-thermal population unambiguously shows that both constituents of each pair originate from the same exciton. We demonstrate that coupled polaron pairs are present even several microseconds after the charge transfer step and suggest that they correspond to Coulomb-bound CTC which are the essential precursor states for the formation of free charge carriers. The dissociation of CTC, and hereby the transformation of the trEPR spectrum to the signatures of separated polarons, can be observed. The resulting polarons (P\textsuperscript{+} in the polymer and P\textsuperscript{-} in the electron accepting material) can be identified by their characteristic $g$-values. Thus, transient EPR measurements enable real-time observation of CTC at the heterojunction and allow us to monitor their dissociation into separated polarons on a timescale ranging from sub-microsecond to several tens of microseconds.

\section{Experimental Details}
The polymer poly(3-hexylthiophene) (P3HT) was purchased from Aldrich and the fullerene [6,6]-phenyl ${\rm C_{61}}$-butyric acid methyl ester (PCBM) from Solenne. No additional purification was performed. Sample preparation took place inside a nitrogen glovebox to avoid exposure to oxygen. The materials were dissolved in the ratio of 1:1 (by weight) in chlorobenzene with a concentration of $15~{\rm mg/ml}$ using magnetic stirring overnight. $50~{\rm \mu l}$ of the solution were loaded into $4~{\rm mm}$ diameter EPR quartz tubes, and the solvent was evaporated under vacuum at $40~{\rm ^\circ C}$, leaving a thick film on the inner sample tube wall. The tubes were subsequently sealed under rough vacuum (final pressure ca.\ $3 \times 10^{-2}~{\rm mbar}$) using a blow torch.

Transient detection of EPR following pulsed laser excitation was performed using a laboratory-built spectrometer.\cite{weber2002_pnas} Optical excitation was provided by a Nd:YAG laser (Spectra Physics GCR-11) pumping an optical parametric oscillator (Opta BBO-355-vis/IR) tuned to a wavelength of $532~{\rm nm}$. The resulting photon energy was close to the absorption maximum of P3HT. The pulse length was $6~{\rm ns}$, and the pulse energy used in the experiments was approximately $5~{\rm mJ}$. Transient EPR signals were recorded by a digital oscilloscope (Tektronix TDS-520A) and accumulated for several pulses for each field position in the scan range of the static magnetic field. The laser-flash-induced background signal was recorded far from the resonant magnetic field and subtracted afterwards from the measurement. The time resolution of our setup is approximately $800~{\rm ns}$ and is primarily limited by the bandwidth of the resonator.

\section{Results and Discussion}
\begin{figure*}
\includegraphics[width=0.70\textwidth]{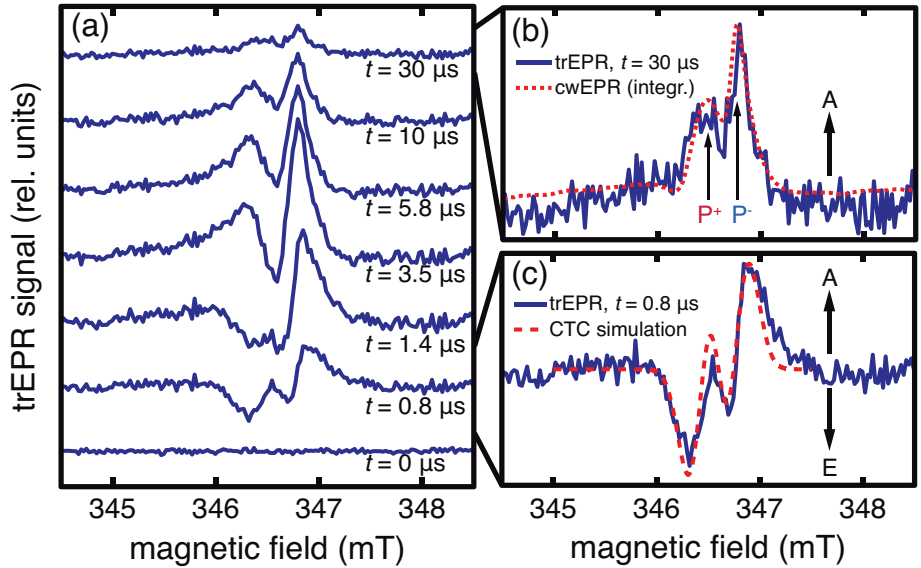}%
\caption{\label{fig_01}(Color online) Series of trEPR spectra of a P3HT:PCBM blend obtained at $T = 100~{\rm K}$. (a) Spectra recorded for several delay times (as indicated) after the laser flash. (b) Zoom into the spectrum at  $t = 30~{\rm \mu s}$ along with the integrated cwEPR spectrum measured under continuous illumination with white light. Both resonance lines have absorptive (A) character. (c) Zoom into the spectrum at  $t = 0.8~{\rm \mu s}$, consisting of absorptive (A) and emissive lines (E), together with the simulation results for a coupled polaron pair. Details are given in the text.}
\end{figure*}

Figure \ref{fig_01}(a) shows a series of trEPR spectra recorded at $T = 100~{\rm K}$ for several delays after the laser flash (at $t = 0$). 
For long delays ($t > 10~{\rm \mu s}$) the spectrum exhibits purely absorptive resonances known from non-interacting separated polarons (P\textsuperscript{+} in P3HT and P\textsuperscript{-} in PCBM, respectively) that result from the dissociation of excitons generated by the laser flash. Figure \ref{fig_01}(b) demonstrates this by comparing the trEPR spectrum at $t = 30~{\rm \mu s}$ and the light-induced cwEPR spectrum (integrated) obtained under white light illumination.

The similarities between the cwEPR and trEPR spectra in terms of line shapes and relative intensities corroborates the interpretation that we indeed observe separated charge carriers in the trEPR spectrum at long delays after the excitation pulse. The loss of charge carriers due to recombination leads to a continuous decay of the free charge carrier trEPR spectrum in the observation time window.\cite{grzegorczyk2010_jphyschemc} 
\subsection{Transient EPR of charge transfer complexes}
In contrast to the signal at $t = 30~{\rm \mu s}$, the spectrum measured at $t = 0.8~{\rm \mu s}$  (see Figure \ref{fig_01}(c)) exhibits both absorptive (A, positive signal) as well as emissive (E, negative signal) components. The presence of emissive lines is a clear indication for coupled geminate PP. The polarization pattern is EAEA and results from the fact that the population of the CTC spin eigenstates differs from the population in thermal equilibrium.\cite{buckley1987_cpl} The photogenerated exciton (cf.\ Figure \ref{fig_02}) is assumed to be in a pure singlet state $|S\rangle$ because of the singlet ground state with all electrons paired in doubly occupied orbitals. Further, the electronic transition induced by the laser flash does not change the spin-multiplicity. Triplet excitons in P3HT may be formed by intersystem crossing and would give rise to characteristic EPR signatures.\cite{liedtke2011_jacs} However, the exciton dissociation time (less than $1~{\rm ps}$) is much shorter than typical intersystem crossing time constants in the range of $100~{\rm ps}$.\cite{cunningham2008_jphyschemc} This is in agreement with the fact that we could not observe triplet excitons here. Due to conservation of angular momentum the resulting CTC inherits the spin-multiplicity of the precursor exciton, i.e.\ only the spin-pair eigenstates with singlet content (mixed states $|2\rangle$ and $|3\rangle$) are populated, whereas the pure triplet states ($|\!\!\uparrow\uparrow\rangle$ and $|\!\!\downarrow\downarrow\rangle$) are not populated. This non-Boltzmann population pattern results in emissive and absorptive lines, which cancel unless the two charge carriers are coupled via exchange or dipolar interaction --- as expected for CTC with a radius of a few ${\rm nm}$. Figure \ref{fig_01}(c) shows the trEPR spectrum at $t = 0.8~{\rm \mu s}$ (close to the time resolution of our setup) along with an easyspin\cite{stoll2006_jmr} simulation assuming the line parameters deduced from the light-induced cwEPR spectrum. Since the spectrum exhibits an almost symmetric EAEA polarization pattern known from paramagnetic centers with isotropic $g$-values and isotropic spin-spin interaction, we include, in a first approach, only isotropic $g$-values ($g = 2.0017$ for P\textsuperscript{+} in P3HT and $g = 1.9998$ for P\textsuperscript{-} in PCBM) as well as an identical line width of $0.31~{\rm mT}$ (FWHM) in the simulation, although both paramagnetic centers have anisotropic $g$-matrices showing rhombic symmetry as evidenced by recent high-frequency pulsed EPR measurements.\cite{poluektov2010_jphyschemb} However, given the disordered nature of the material suggesting that there is no fixed relative orientation between both $g$-matrices, we believe that these simplifications do not qualitatively impair the simulation results. Note that the line widths used here are considerably larger than those found in cwEPR spectra of separated polarons in P3HT and PCBM. We further assume a non-thermal polarization (singlet precursor state). To account for the spin-spin interaction between P\textsuperscript{+} and P\textsuperscript{-}, we consider the simplest possible situation and include an isotropic exchange interaction ($J = 1~{\rm MHz}$) only. Dipolar coupling (quantified by the dipolar interaction strength $D$) is certainly present as well and may be of comparable size or even larger than $J$. Yet, extracting reliable values for $J$ and $D$ is impossible for relatively broad EPR lines. While they critically determine the polarization pattern (i.e. EAEA or AEAE), the general shape of the trEPR spectrum is rather insensitive to the exact values of $J$ and $D$ for sufficiently small couplings. For this reason we restrict our description to the exemplary case of a purely isotropic exchange interaction. Even these rough approximations lead to an excellent agreement between simulation and experimental results (cf.\ Figure \ref{fig_01}(c)). We can draw two important conclusions from these observations:

1.\ Coupled polaron pairs exist even several microseconds after exciton generation and subsequent charge transfer at $T = 100~{\rm K}$. The fact that we observe a non-Boltzmann polarization pattern, which results from the precursor singlet exciton, unambiguously demonstrates that the PP constituents originate from the same exciton, i.e.\ they are geminate in their origin. The presence of spin-spin coupling necessitates both charge carriers to reside in close proximity and hence to experience Coulomb attraction. The $g$-values show that the pairs are formed at the donor/acceptor interface. We thus argue that the coupled charge carriers observed in the trEPR spectrum embody the charge transfer complexes mediating the dissociation of excitons into separated polarons, which were previously detected by various optical techniques.\linebreak
2.\ CTC are involved in free polaron generation as evidenced by the good agreement between the trEPR spectrum at long delays and the light-induced cwEPR spectrum.

\subsection{From CTC to separated polarons}
\begin{figure}
\includegraphics{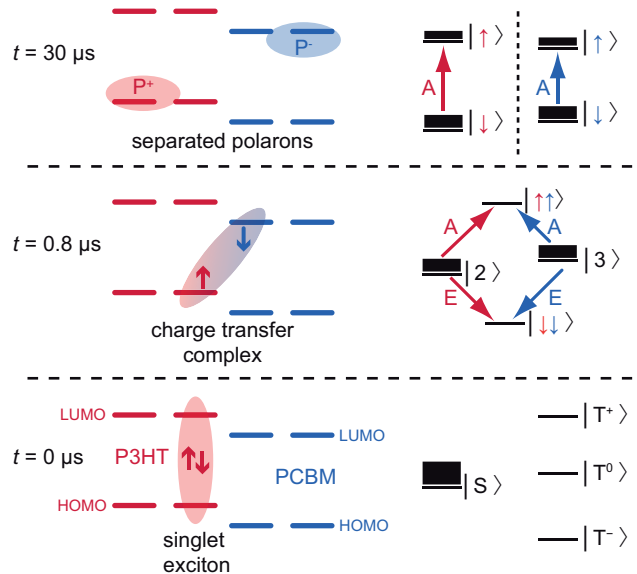}%
\caption{\label{fig_02}(Color online) Schematic energy diagram of the HOMO and LUMO levels of P3HT and PCBM (left) and corresponding spin-pair eigenstates of a charge transfer complex consisting of two $S = 1/2$ particles (right) along with the EPR-active transitions (as indicated by the arrows) which give rise to absorptive (A) and emissive (E) signals in the trEPR spectrum. From bottom to top a singlet exciton ($|S\rangle$) is created upon photoexcitation of the P3HT. This is subsequently converted into a CTC, with the mixed states $|2\rangle$ and $|3\rangle$ being populated and the pure triplet states $|\!\!\uparrow\uparrow\rangle$ and $|\!\!\downarrow\downarrow\rangle$ being unpopulated (as indicated by the according horizontal bars). Finally, the CTC dissociates into two separated polarons (P\textsuperscript{+} and P\textsuperscript{-}) exhibiting thermalized spin populations.}
\end{figure}

The change of the trEPR spectrum resulting from the transition of the coupled pair ($0.8~{\rm \mu s}$) towards free polarons ($30~{\rm \mu s}$), which is shown schematically in Figure \ref{fig_02}, is not fully understood yet. 
Several mechanisms can give rise to a temporal variation of the polarization pattern, among them selective depopulation of the individual spin states at different rates. Possible underlying mechanisms include mixing processes, which provide coupling (accompanied by polarization transfer) between the constituents of the pairs,\cite{adrian1979_jchemphys} or (orientation-) selective recombination.\cite{shushin1997_cpl} These processes may also affect the spectrum attributed to separated polarons. Transient spin nutations can also give rise to a polarization pattern that varies with time.\cite{gierer1991_cpl} In contrast to radical pairs in most biological systems, the situation in polymer:fullerene blends is further exacerbated by the fact that the PP radius is not fixed, but is rather a blend-morphology driven distribution and also may vary as a function of time after the laser flash. This effect is accompanied by a temporal change in the spin-spin interaction parameters and, in turn, additionally contributes to spectral changes. In addition, the amplitude of the free polaron spectrum, which superimposes the spin-correlated PP spectrum, gradually decays due to charge carrier recombination.

The strength of the spin-spin coupling, which manifests in the spectrum at short delays, in principle allows determining the CTC radius $r$.\cite{stehlik1989_jphyschem} Since the dipolar interaction strength varies with $r^{-3}$, measurement of $D$ directly determines $r$, provided that both spins can be treated as point dipoles. The so-called point-dipole approximation is justified only as long as the spatial distribution of the wavefunction of each spin is much smaller than $r$, which is true for most radical pairs in biological systems --- the main application field for trEPR so far. In our case the polymer polaron is delocalized over several monomer units,\cite{brendel1991_synthmet,zezin2004_cpl,chang2006_prb} making the point-dipole approximation not applicable here. In addition, quantification of $D$ from trEPR spectra is generally complicated for small spin-spin interactions in the presence of broad lines that result from substantial hyperfine coupling of the unpaired spins to magnetic nuclei. We do not attempt at extracting the CTC radius here but restrict our conclusion to the fact that even rough approximations (isotropic $g$-values, isotropic exchange interaction) based on cwEPR results can qualitatively reproduce the measured spectrum. Extracting a reliable value for the pair radius would require considering the exact wavefunction of both polarons, which is feasible but beyond the scope of the present article. We note that an EPR-based technique called out-of-phase electron spin echo envelope modulation (OOP-ESEEM) is available that is specifically sensitive to interactions in spin-correlated pair systems.\cite{salikhov1992_amr} OOP-ESEEM was previously used to determine the dipolar interaction strength within radical pairs in photosynthetic reaction centers\cite{bittl2001_bba} and recently also applied to donor-acceptor systems for artificial photosynthesis.\cite{carmieli2009_jacs}

Since the signal intensities in trEPR cannot be quantified directly, i.e.\ we do not exactly know how many CTC or free polarons contribute to the spectrum, we cannot \emph{a priori} exclude that the detected long-lived CTC represent only a minor fraction without substantial relevance for free charge carrier generation. However, we can confirm that the CTC that show up in the trEPR spectrum indeed represent a relevant species based on plausibility arguments: It is generally accepted that EPR is sensitive to free polarons which are responsible for the electrical conductivity in organic semiconductors. This holds true not only for light-induced cwEPR, but also for trEPR (cf.\ Figure \ref{fig_01}(b)). If only a small fraction of excitons formed CTC after charge transfer, whereas the dominant pathway immediately led to free polarons via a ``hot process'',\cite {lee2010_jacs} we would expect to detect these free charge carriers in the trEPR spectrum as well. Even if the free polarons decayed quickly on the observation timescale, the remaining signal intensity at $t = 30~{\rm \mu s}$ suggests that the trEPR spectrum would be dominated by free polarons (exhibiting a thermalized spin population) at short delay times. The fact that the contribution from free polarons to the spectrum at $t = 0.8~{\rm \mu s}$ is negligible, leads us to the conclusion that the dominant pathway towards free charge carriers occurs via CTC.

It is expected and confirmed by preliminary measurements that the temperature has a significant influence on the CTC dynamics. In consequence, the dissociation time of the CTC under solar cell operating conditions will certainly deviate from the dissociation time at $T = 100~{\rm K}$. However, as pointed out before, the charge separation yield in P3HT:PCBM blends was reported to be almost independent of temperature\cite{grzegorczyk2010_jphyschemc} and applied electric field,\cite{deibel2009_prl} rendering our observations significant for solar cell conditions, despite the potentially longer CTC dissociation time at $T = 100~{\rm K}$. Studies on the temperature behaviour of CTC dynamics are currently being performed.

We note that in our experiment the concentrations of excitons, CTC and free carriers generated by the intense laser flash are not representative of the operating conditions of organic solar cells. A detailed analysis of the trEPR signals as a function of the light intensity and the bias voltage applied to fully processed devices will allow us to evaluate the relevance of CTC, as detected by trEPR, for solar cell operation. Furthermore, trEPR studies on blends that experienced different post-treatments may provide helpful insight into the crucial influence of the blend morphology\cite{yang2007_macromolecules} on charge carrier separation in polymer:fullerene solar cells.

The model system P3HT:PCBM was chosen for this exemplary study because of its well known electrical and optical properties. Extending trEPR investigations to low-bandgap copolymers and small molecules used in high efficiency organic solar cells will offer the possibility to elucidate the role of CTC in these important classes of materials. The characteristic $g$-value of P\textsuperscript{-} in PCBM, which significantly differs from the $g$-values of P\textsuperscript{+} in most materials used as electron donor, will provide sufficient spectral resolution to enable trEPR measurements on a wide variety of materials when blended with PCBM even at relatively low resonance frequencies.

\section{Conclusion}
This study clearly demonstrates that trEPR is a valuable tool to characterize CTC, complementing time-resolved optical measurements. The low-temperature trEPR spectrum of a polymer:fullerene blend immediately following photoexcitation of P3HT and subsequent charge transfer to PCBM reveals signatures of coupled polaron pairs. The pairs inherit the spin-multiplicity of the singlet exciton and thus lead to spin-correlation, resulting in a trEPR spectrum that reveals emissive and absorptive components. Since the PP spin states exhibit non-thermal population, we can attribute the trEPR signals to geminate polaron pairs originating from one exciton. The $g$-values associated with the contributing paramagnetic centers show that the pairs are located at the donor/acceptor interface. In consequence, we suggest that the spin-correlated PP observed in the trEPR spectrum correspond to Coulomb-bound CTC mediating the dissociation of excitons into separated charge carriers, which were previously detected by various optical techniques. At $T = 100~{\rm K}$ CTC can be observed even for delays $t>1~{\rm \mu s}$ after the laser flash. In contrast, for long delays ($t>10~{\rm \mu s}$) the spectrum consists of purely absorptive resonances known from non-interacting polarons that are typically detected using light-induced cwEPR.

Upon increasing temperature the CTC dissociation becomes faster and eventually exceeds the time resolution of our setup at room temperature. However, trEPR measurements with time resolutions down to $1~{\rm ns}$ and below are conceivable\cite{vantol2005_revsciinstr} and will certainly be beneficial when extending the analysis towards temperatures closer to ambient conditions.

The fact that coupled polaron pairs as well as separated polarons can be detected in the same measurement exemplifies that the trEPR technique allows studying the dynamics of intermediate species involved in charge transfer and charge separation in materials relevant to organic solar cells.

\begin{acknowledgments}
We thank Carsten Deibel for helpful discussions. Freie Universit\"at Berlin and Helmholtz-Zentrum Berlin acknowledge financial support from BMBF (Grant No.\ 03SF0328). The work at the Universit\"at W\"urzburg was supported by the DFG within the DY18/6-1, 18/6-2 and INST 93/557-1 FUGG projects.
\end{acknowledgments}
\bibliography{bib_transient}

\end{document}